\documentclass[num-refs]{wiley-article}

\usepackage{graphicx}

\usepackage{siunitx}

\usepackage[caption=false,font=footnotesize]{subfig}

\papertype{Original Article}
\paperfield{Journal Section}

\title{Mining the Relationship Between COVID-19 Sentiment and Market Performance}

\author[1]{Ziyuan Xia}
\author[2]{Jeffrey Chen}
\author[3]{Anchen Sun}

\affil[1]{Antai College of Economics \& Management, Shanghai Jiao Tong University, 1954 Huashan Road, Shanghai, China}
\affil[2]{Decision, Operations \& Information Technologies, University of Maryland Robert H.Smith School of Business, 7621 Mowatt Ln, College Park, MD 20742, U.S.}
\affil[3]{Department of Electrical and Computer Engineering, University of Miami, Coral Gables, U.S.}

\corraddress{Ziyuan Xia, Antai College of Economics \& Management, Shanghai Jiao Tong University, 1954 Huashan Road, Shanghai, China}
\corremail{edwardxzy@sjtu.edu.cn}

\fundinginfo{Ziyuan Xia}

\runningauthor{Ziyuan Xia et~al.}

\begin{document}

\begin{frontmatter}
\maketitle

\begin{abstract}
At the beginning of the COVID-19 outbreak in March, we observed one of the largest stock market crashes in history. Within the months following this, a volatile bullish climb back to pre-pandemic performances and higher. In this paper we study the stock market behavior during the initial few months of the COVID-19 pandemic in relation to COVID-19 sentiment. Using text sentiment analysis of Twitter data, we look at tweets that contain key words in relation to the COVID-19 pandemic and the sentiment of the tweet to understand whether sentiment can be used as an indicator for stock market performance. There has been previous research done on applying natural language processing and text sentiment analysis to understand the stock market performance, given how prevalent the impact of COVID-19 is to the economy, we want to further the application of these techniques to understand the relationship that COVID-19 has with stock market performance. Our findings show that there is a strong relationship to COVID-19 sentiment derived from tweets that could be used to predict stock market performance in the future.

\end{abstract}
\end{frontmatter}

\section{Introduction}
Understanding and predicting the stock market behavior has always been a major goal for academia and industry, but an extremely difficult one. The efficient market hypothesis, which is an earlier theory of stock market prices by economist Eugene Fama, states that in an efficient market the current market prices is reflective upon all available and relevant information~\cite{fama1965behavior, fama1969adjustment}. Any changes in stock market prices are a result of new information that is revealed, independent of existing information. As such, it was believed that the market price was unpredictable because new information is also unpredictable. Weaker forms of market efficiency are consequences of incomplete information to the public, causing the true value of a stock price to not be represented accurately.
When information asymmetry is reduced, we see market anomalies and volatility reduce or disappear. By increasing the access to information and transparency to the public, strong market efficiency is achieved~\cite{fama1991efficient}. 

In theory, strong market efficiency is achieved when we have perfect information. However, easily accessible information is not always readily available to the general public, and when it is, there is delay from the actual event to news reporting. Although news can be unpredictable, there are certain indicators that can be derived from online social media platforms and tools that could act as form of asymmetric new information to predict stock market performances. 

In this paper, we want to understand how public sentiment from social media platforms can be used to determine stock market performance. More specifically, how the public sentiment of the severity of the COVID-19 pandemic with regards to how it is being handled, infection rates, and confidence in economy and government, affects consumer investment behavior and as a result, stock market performance~\cite{stojkoski2020socio}.

The reaction of the major global equity markets to COVID-19 in the early 2020 can be basically divided into the following parts in terms of timing~\cite{CHERNOZHUKOV202123}. 
\begin{itemize}
    \item Early stage of the crisis: Infection is limited to Asia, with only the Hong Kong stock market falling more significantly. 11 January 2020, the first fatal case of New Coronary Pneumonia is reported in Wuhan, China, confirming the human-to-human nature of the virus. Wuhan was closed on January 23, and confirmed cases continued to emerge in Thailand, Japan, Korea, Hong Kong, and Singapore. The World Health Organization announced on January 31 that it had upgraded the NCCP outbreak to an "international public health emergency", followed by the Diamond Princess incident in Japan and the declaration of community transmission in Korea, etc~\cite{liu2020response}. However, although the global market was concerned about the resumption of work in China and the outbreak in Asia, the confirmed cases were still mainly in Asia, so the stock markets of various countries did not yet reflect the NCCP outbreak significantly. However, as the confirmed cases are still mainly in Asia, the global stock markets have not yet reflected the NCCP epidemic significantly, with only small gains and losses~\cite{zhang2020financial}.
    \item In late February, a massive outbreak of Newcastle pneumonia occurred in Europe, with the number of newly diagnosed patients rising; after the partial closure of the Italian Lombardy region on March 4, European and global stock markets took a sharp turn for the worse. With the news that the healthcare system in Europe and the U.S. was on the verge of a tipping point in March, global stock markets fell with violent shocks, and U.S. stocks triggered four market meltdowns between mid-March and the end of March~\cite{huang2020covid}. The Dow Jones collapsed $2,352$ points on March 12 (down $10\%$ on that day) and fell another $2,997$ points on March 17 (down $13\%$ on that day), the largest drop in history, triggering global panic~\cite{zeren2020impact}.
\end{itemize}
From January to Mid-March 2020, European and U.S. stocks fell by a deep $-24.9\%$ to $-28.6\%$, while Asian markets such as Japan, Korea and New Zealand fell by a relatively small amount ($-18.7\%$ to $-22.3\%$). As the country where the outbreak originated, Shanghai and Hong Kong stocks fell only $-10.4\%$ and $-18.0\%$, respectively, due to the significant decrease in new cases announced since March, making them the rare markets that resisted the global stock market meltdown. 
By April 2020, though infection rates have not improved~\cite{manski2020estimating}, the stock market has been in steady recovery, indicating that the news of growing coronavirus cases was no longer negatively impacting the stock market~\cite{topcu2020impact, salisu2020predicting}. As of June 8, 2020, the WHO had announced that the COVID-19 situation is still worsening, but at the same time, the stock market was still growing for the fourth week in a row, with the S\&P 500 returning to it's position before the pandemic~\cite{Thestock98:online}.

The pandemic's impact on the economy~\cite{KEANE202186}, and public opinion and confidence in the government and institutions responsible for handling the pandemic, with regards to areas in supply chain, leadership, policy and research, can have a high impact on consumer behavior in investing and spending. A study conducted on consumer panic on over 54 countries using the first fourth months of the 2020 using Google search data showed that consumer panic occurred at various levels of severity over the course of the four months, with some cases happening earlier, and some later. This study showed us how the severity of consumer panic can occurs when a government establishes movement restrictions  (Consumer panic in the COVID-19 Pandemic, Keane and Neal). Another study done on the policies of mandating face masks, stay-at-home orders, and restricting non-essential businesses have effectively reduced the spread of Covid-19. (Causal Impact of Masks, Policies, Behavior on Early Covid-19 Pandemic in the US, Chernozhukov, Kasahara). Consumer perception of an economic recovery or government's competency can be caused through the enactment of government policy and its impact on supply chain. This new information could then be reflected in the value of the a stock during specific economic and political times. Other research has found that for positive and negative COVID-19 information has a heavy impact on market volatility~\cite{baek2020covid}. 
Therefore we want to capture the population sentiment with an indiscriminate sample of new information that could reflect the public perception of the COVID-19 situation. The method we proposed is natural language processing for text sentiment analysis.

We found Twitter to be an ideal source of capturing natural language public sentiment data, as it is an social media platform with over 300 million monthly active users. 
Early 2010 research of understanding behavioral economics from a societal mood states indicates that it is possible to use text sentiment using deep learning models to correlate with the performance of the Dow Jones Industrial Average (DJIA) over time~\cite{bollen2011twitter}.  Later research found that in a short-window event study of a UK based political event, Twitter messages (tweets) were collected and filtered to labels pertaining to a political event. With later statistical forecast proving evidence of causation between the public sentiment, and the closing price with a slight time lag~\cite{nisar2018twitter}. We further the research of using text sentiment analysis in order to assess whether public opinion representative of a consumer's confidence in the COVID-19 situation could also be used to predict the stock market indexes~\cite{baker2020unprecedented}. 

To conduct our research, we used data extracted from the Twitter API~\cite{TwitterA46:online}. The technology affordance of social media platforms like Twitter allow us to filter the text data from Twitter that contains keywords which indicate that the message is pertaining to COVID-19. We then record the sentiment score, normalize the scores, and compare it to stock market index performance over time. For the stock market indexes, we used the historical data of the S\&P 500, DJIA, and NASDAQ that was retrieved using Yahoo Finance API~\cite{YahooFin20:online}.

\section{Related Work}

Predicting stock prices is nothing new. In the field of econometrics, many different methods have been applied to the prediction of stock prices. One of the famous model is Financial Time Series (FTS)~\cite{tsay2005analysis}. FTS modeling has a long history, having first revolutionized algorithmic trading in the early 1970's. FTS analysis consists of two types of analysis: fundamental and technical. However, both types of analysis have been challenged by the efficient market hypothesis (EMH)~\cite{malkiel2003efficient}, a controversial hypothesis that has been around since 1970.

Since its introduction in 1970, the EMH has been controversial, assuming that stock prices are ultimately unpredictable. This does not limit the study to FTS modeling by using linear, nonlinear, and ML-based models. Because financial time series are non-stationary, non-linear, and noisy, it is difficult for traditional statistical models to predict them accurately. In recent years, more and more studies have attempted to apply deep learning to stock market forecasting, although it is still far from perfect.

In~\cite{lin2013svm} propose a support vector machine (SVM) based stock prediction method to develop a two-part feature selection and prediction model, and demonstrate that the method has better generalization ability than traditional methods. In~\cite{wanjawa2014ann} propose a neural network for predicting stock prices using a feedforward multilayer perceptron with backpropagation of errors. The results show that the model is capable of predicting a typical stock market.

The research entered the LSTM era in 2017 and the proliferation of research using LSTM networks to process time series data. LSTM was proposed by~\cite{hochreiter1997long} and recently refined and popularized by Alex Graves. \cite{zhao2017time} propose to add a time-weighted function to LSTM and the results outperform other models.

\cite{qiu2020forecasting} It combines the LSTM and an attention mechanism to design an attention-based LSTM then compares it with the LSTM model, the LSTM model with wavelet denoising, and the gated recurrent unit(GRU) neural network model to show the advantages of the incorporation with the attention mechanism. Around the same time, a new architecture of neural network, Deep Wide Area Neural Network (DWNN), is proposed. The results show that the DWNN model can reduce the mean squared error of the forecast by 30\% compared to the conventional RNN model. \cite{kim2018forecasting} proposed to integrate CNN and DWNN models into a single model, which can reduce the mean-squared error of forecasts by 30\% compared to conventional RNN models. A hybrid neural network model is proposed for a quantitative stock selection strategy to determine stock market trends and then to predict stock prices using LSTM, and a hybrid neural network model is proposed for a quantitative timing strategy to increase profits. In \cite{jiang2018stock,wu2020graph} use LSTM neural network, graph network and RNN to build models and find that LSTM can be better applied to stock prediction. In their paper~\cite{jin2019stock} added investors' sentiment propensity to the model analysis and introduced empirical modal decomposition (EMD) in combination with LSTM to obtain more accurate Stock Prediction. LSTM models based on attentional mechanisms are common in speech and image recognition, but are rarely used in finance.

For financial markets, deep learning methods cannot be applied directly to the stock market. Specifically, the ability of algorithms such as LSTM to handle serial data has been proven in scientific research over a long period of time. However, for stock analysis, the predictive power of these algorithms is far from adequate. The stock market is not at all as simple as the analysis of serial data. There have been thousands of projects in the scientific field trying to use LSTM or other time series analysis methods to predict stocks, but there are few published algorithms that can be practically applied in the market. The so-called deep learning is just an inductive method based on fitting historical data. If deep learning is used to make stock predictions, the long-term returns will definitely be negative because the market is changing, the laws are changing, and history may repeat itself but it will not be the same.

It is not meaningless to apply deep learning to the stock market. For example, sentiment analysis can be applied to news, social media, etc. to analyze the overall sentiment of the market towards a particular stock or a particular sector~\cite{haroon2020covid,cao2019stock,shi2021stock}. 
The stock market is essentially a game process, and observing the game process through sentiment analysis is a better entry point for introducing existing analytic algorithms into the financial market. Stock prediction is never a matter of putting a simple time series data into deep learning and making money. There is a wide variety of data on stock buy points, trading volume, historical prices, etc., and they serve different purposes. Instead of trying to uncover the complex mathematical models of the stock financial market, we should change the entry point and analyze the relationship between stock price changes and emotional fluctuations from the perspective of the emotions of ordinary stockholders who buy stocks~\cite{schumaker2009textual}. 

\section{Data}
In this section, we will describe how the data used in our study was obtained. By building a sentiment analysis tool, we obtained a fast, efficient and autonomous way to obtain real-time data and store it on a server. The Twitter data will help us to monitor the sentiment of Twitter users towards COVID-19 and can be generalized to any other topic after more experiments. In the stock market, we use Yahoo Finance's API to obtain data on the US stock market.

\begin{table}
\centering
\caption{Keywords used to crawling COVID-19 Tweets}
\begin{tabular}{lllllll}
\hline
Epidemic   &              &               &            &          &          &                                   \\ \hline
covid-19   & corona       & virus         & pandemic & mask     & stay home    & work from home    \\
breathing  & China        & Wuhan         & lock down  & outbreak & testing site & asymptonmatic               \\
quarantine & vaccine & CDC   & N95      & KN95     & transmission      & community spread \\ 
endemic & epidemic & flu shot &  positive & sars-cov-2      &  isolation       &  \\ \hline
Panic-Buying   &              &               &                  &          &              &       \\ \hline
toilet paper & pasta & rice & hoarding & fruit & vegetables & panic buying \\
supermarket & flour &  &  &  &  &  \\\hline
\end{tabular}
\label{tb:keywords}
\end{table}

\subsection{Data Collection and Sentiment Analysis}
We built a sentiment analysis tool that extracts COVID-19 related tweets every thirty minutes. A sentiment analysis model trained on the Twitter data was used to analyze these stored tweets and output a sentiment score every 30 minutes. The sentiment scores are defined in the range $[-1,0)$, $0$, and $(0,+1]$ for negative sentiment, neutral sentiment, and positive sentiment. 

We created a cluster on the Google Cloud Platform (GCP)~\cite{Financia31:online} and built a Twitter text data streaming sentiment analysis tool with Flume, PySpark, and PyTorch. Twitter offers developer accounts, which we use to access Twitter's API and set up the Flume Collecter on GCP cluster to collect streaming tweets. The keywords are shown in Fig.~\ref{tb:keywords} set to fetch tweets from API. 

The tool reads the Tweets streaming from Hadoop Distributed File System (HDFS) first, which can scan the HDFS path and convert all readable text files in the path to RDD format. After we get the completed streaming data file, we need to clean the text file because Flume streams all texts given by Twitter API. After reviewing the raw data text file, we found that these texts were encoded by Unicode and contained different international languages. The texts also includes emojis, Apple Emoticon Package, href and HTTP links etc. The regular expression is the most popular way to clean text data. A regular expression, regex or regexp is a sequence of characters that define a search pattern. We also built a Unicode range check to filter out Chinese, Korean and Japanese. When people write a sentence in these languages, they will not include a space between every two words. The best way to split these sentences is by using a Nature Language Package to mark the sentences then split them. We simply split characters in Chinese, Korean and Japanese and keep other languages as the original. After the initial cleaning of the metadata, we stored these tweets on the server and process our pre-trained PyTorch sentiment analysis model to generate the sentiment score for the tweets collected per thirty minutes intervals.

\subsection{Stock Price}
On the server, we ran the Yahoo Finance API and recorded the daily stock market data of the daily opening and closing prices, as well as trading volume for the three major stock indices (NASDAQ Composite, DJIA, and S\&P 500). The stock market is not open on weekends and holidays, but still has trades occurring. Therefore we use Lagrangian interpolation to fill in all the missing data so that it corresponds to the daily sentiment analysis scores.

Financial time series data (especially stock prices) are subject to constant fluctuations due to seasonality, noise, and automatic corrections~\cite{jiang2020time}. Traditional forecasting methods (FTS) use moving averages and differentials to reduce forecast noise. However, FTS is often unstable and there is overlap between useful signals and noise, which makes traditional denoising methods ineffective.

Wavelet analysis has made impressive achievements in the fields of image and signal processing. It can compensate for the shortcomings of Fourier analysis, and is therefore gradually being introduced into the economic and financial fields. The wavelet transform has a unique advantage in solving traditional time series analysis problems because it can decompose and reconstruct financial time series data from a wide range of time and frequency domains. The wavelet transformation essentially uses multi-scale features to denoise the data set, thus separating the useful signal from the noise efficiently~\cite{qiu2020forecasting} used the coif3 wavelet function for three decomposition layers and evaluated the effect of the wavelet transform by the signal-to-noise ratio (SNR) and root mean square error (RMSE).The higher the SNR, the smaller the RMSE, and the higher the SNR, the smaller the RMSE, and the lower the denoising of the wavelet transform, the better the results. By using the following equation 
\begin{equation}
    \text{SNR} = 10\times \log \left [\frac{\sum_{j=1}^Nx_j^2}{\sum_{j=1}^N(x_j-\hat{x}_j)^2} \right ]
\end{equation} 
we can denoise the collected stock data, we can get the cleaned data for our method.

\subsection{Normalization and Comparison}

\begin{figure}
    \centering
    \includegraphics[width=1\textwidth]{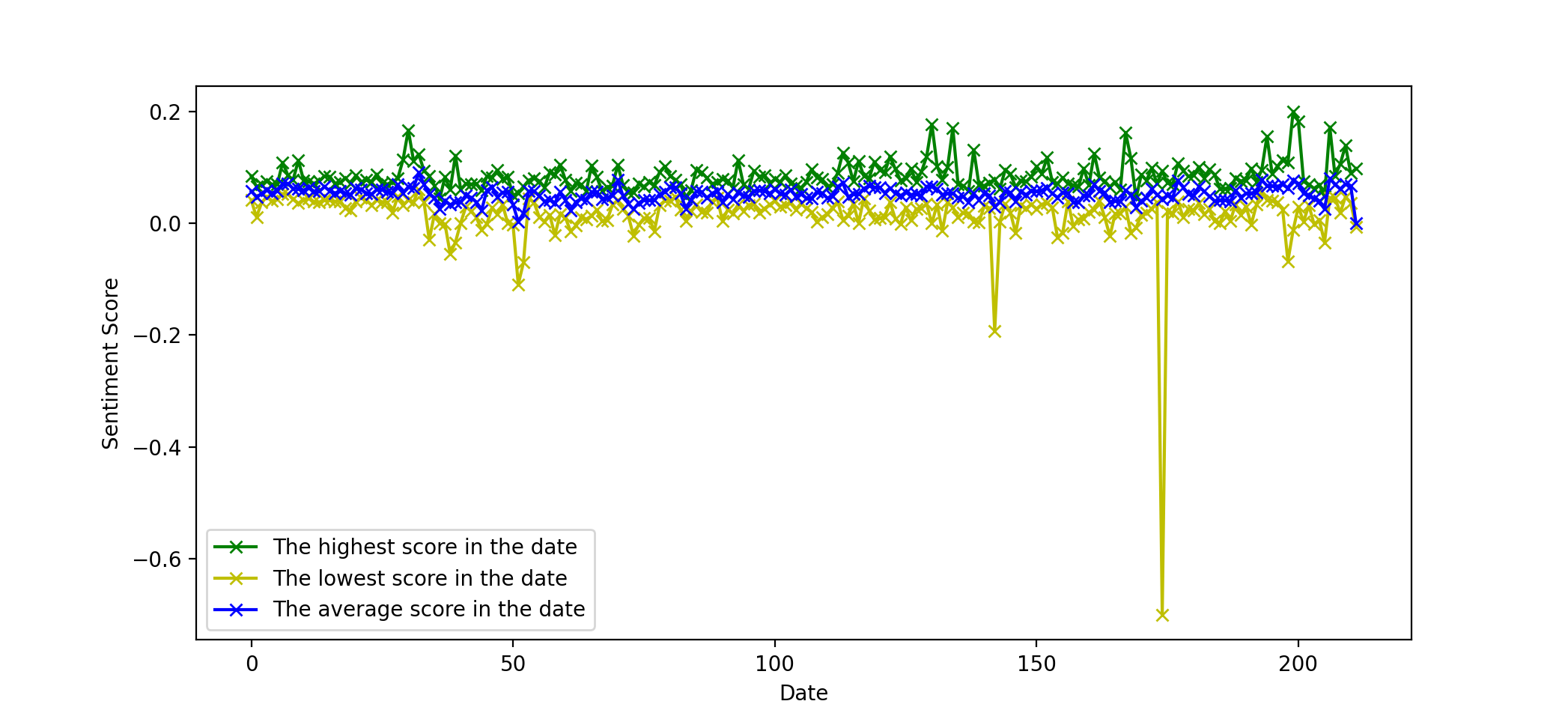}
    \caption{Sentiment Score for COVID-19 Tweets}
    \label{fig:s_score_covid}
\end{figure}

\begin{figure}
    \captionsetup[subfloat]{justification=centering}
    \centering
    \subfloat[Dow Jones Industrial Average]{\includegraphics[width=1\textwidth]{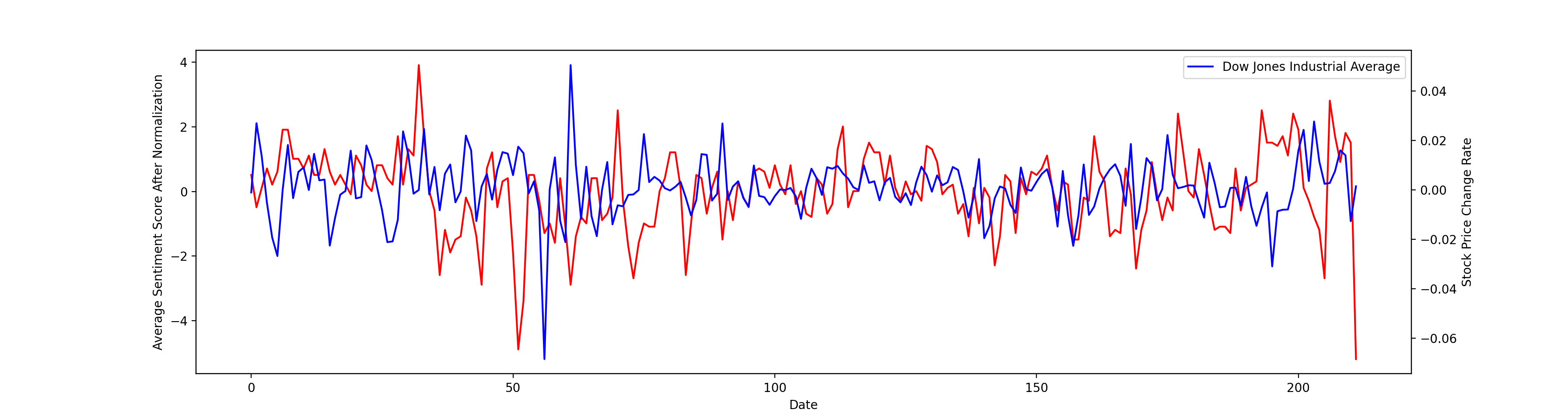}} \\
    \subfloat[Nasdaq Index]{\includegraphics[width=1\textwidth]{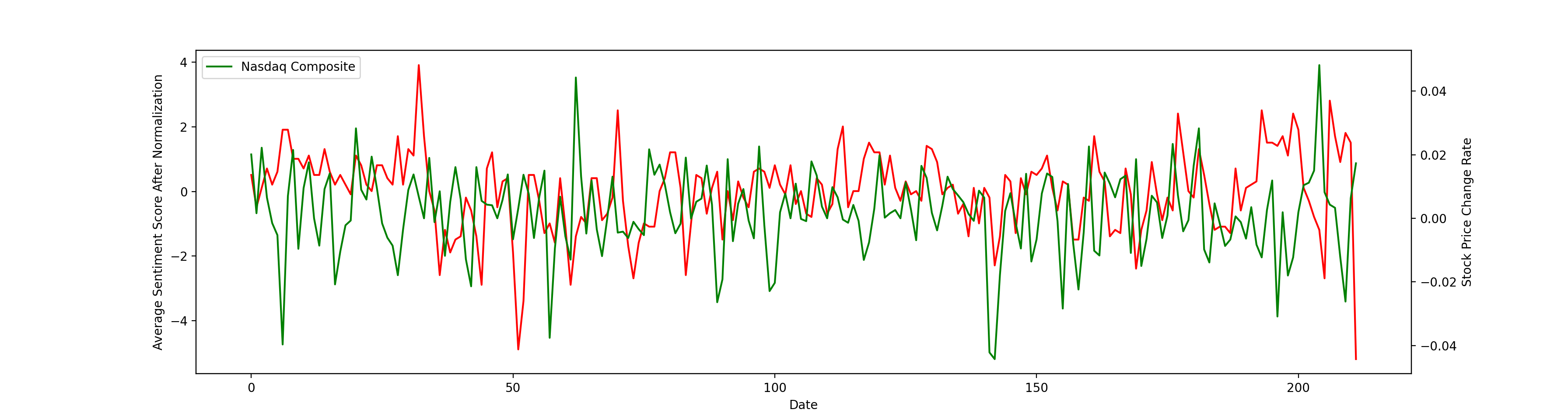}} \\
    \subfloat[Standard and Poor's (S\&P) 500]{\includegraphics[width=1\textwidth]{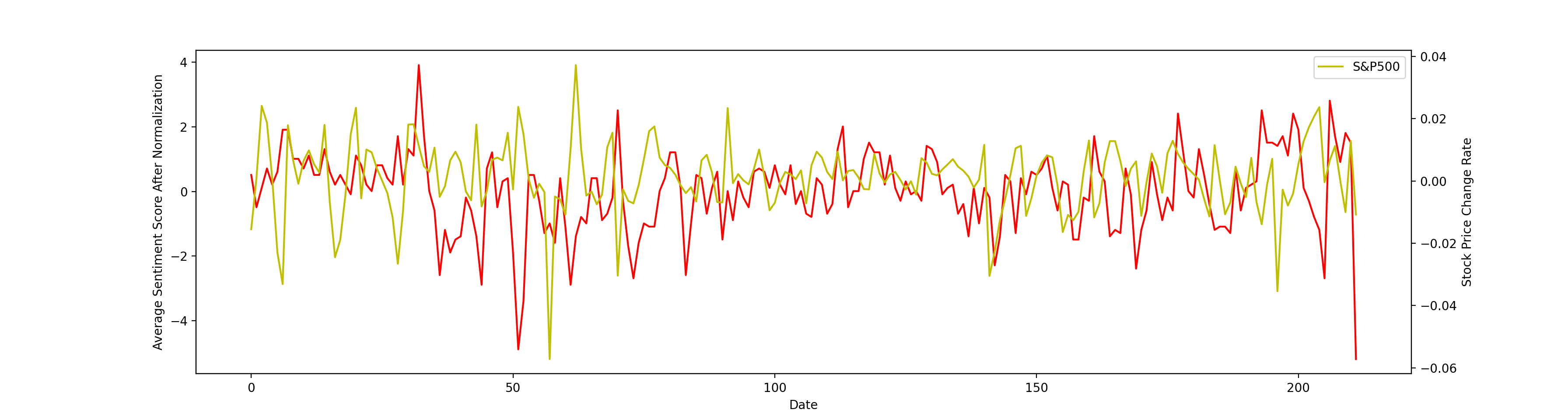}}
    \caption{Comparison of fluctuations in the stock index and regularized sentiment scores during COVID-19}
    \label{fig:stock_index}
\end{figure}

In the sentiment analysis, we found that based on existing models trained on Twitter data, the average sentiment score for 98.1\% of the dates was greater than 0 (positive), which could be due to the fact that the majority of Twitter users have a positive view of the COVID-19 situation, or due to the fact that the models used in our analysis were trained on regular Twitter data, which makes the models more likely to be positive than negative in the analysis. When tweeting about COVID-19, the results were somewhat on the positive side. Based on this perception, and to justify the study, we normalized the obtained sentiment scores by $S_{i} = S_{i} - \bar{S}$ and compared them to the three major stock indices, and the results are shown in Fig. \ref{fig:stock_index}.

\subsection{Analysis}
Based on a comparison of sentiment scores and stock market changes, we can perform a preliminary analysis on the effect of Twitter-related changes in COVID-19 sentiment on stock prices. 

The Dow Jones Industrial Average has often been a key indicator of the overall market performance for the U.S. stock market. As the oldest stock price index in the world with a history of over 100 years, it is comprised of only 30 constituent corporations, of which, are the 30 largest and most well-known listed companies in the United States. However, with more than 10,000 stocks listed on the U.S. stock market, many experts and scholars doubt the capability of the DJIA to be an effective market index, with their 30 constituents. Still, we should note that the 30 constituents are all significant corporations in the United States, each with a large reference value that could be used as an indicator of the overall market performance for an investors reference. We can also find from the generated sentiment analysis that the Dow Jones is the one that is most closely correlated with the change in sentiment scores.

The NASDAQ index was created in 1971 as a key indicator of technology stocks around the world. The constituents include all shares listed on the NASDAQ in the United States and is a key indicator of technology stocks around the world. The NASDAQ index has more than 5,000 constituents, covering all aspects of biotechnology, such as computer hardware, software, semiconductors, network communications, etc. It is the preferred reference for investing in technology stocks. But relatively, the changes in the index of technology stocks relative to the changes in sentiment scores relatively large differences, which may be because technology stocks and the company's own technology level and product technology research and development more relevant, less affected by the COVID-19 situation of public sentiment.

The S\&P 500 Index is an overall measure of the top 500 publicly traded companies in the U.S. The rating company Standard \& Poor's has selected 500 leading companies in various industries (and the 500 largest companies in the U.S. with the highest market capitalization) in the U.S. stock market based on market capitalization and liquidity, selected to cover the two major U.S. stock exchanges (New York Stock Exchange and Nasdaq Stock Exchange). The S\&P 500 contains more companies than the Dow Jones Industrial Average, and therefore better reflects changes in the stock market and is more risk diversified. In addition, the S\&P 500 and the Dow Jones Industrials use different weightings, with the Dow being weighted by stock price and the S\&P 500 being weighted by market capitalization, which better reflects the actual value of a company's stock and can even reflect the rise and fall of the U.S. economy.

\section{The Proposed Method}

In 2020, one of the most important events in the world is COVID-19, a global epidemic that has affected all industries and caused extreme volatility in stock market prices. Due to COVID-19 and its impact on work life, we believe that starting with people's sentiment towards COVID-19 is an excellent attempt to analyze stock prices by exposing the impact of the public's sentiment towards the pandemic on social media with their investments and information on the financial sector in such an unprecedented pandemic. After the COVID-19 outbreak, many researchers turned their attention to social media and tried to uncover useful information related to COVID-19. Nowadays, Twitter is considered one of the reliable indicators for analyzing the spread of epidemics, and the data generated by users' activities on social media is becoming one of the important bases for discovering ways to track and analyze epidemic outbreaks~\cite{polyzos2020tourism}. Thus, we use both time series stock index and COVID-19 tweets sentiment analysis score as input data for the proposed Sentiment-LSTM model. 

\begin{figure}
    \centering
    \includegraphics[width=1\textwidth]{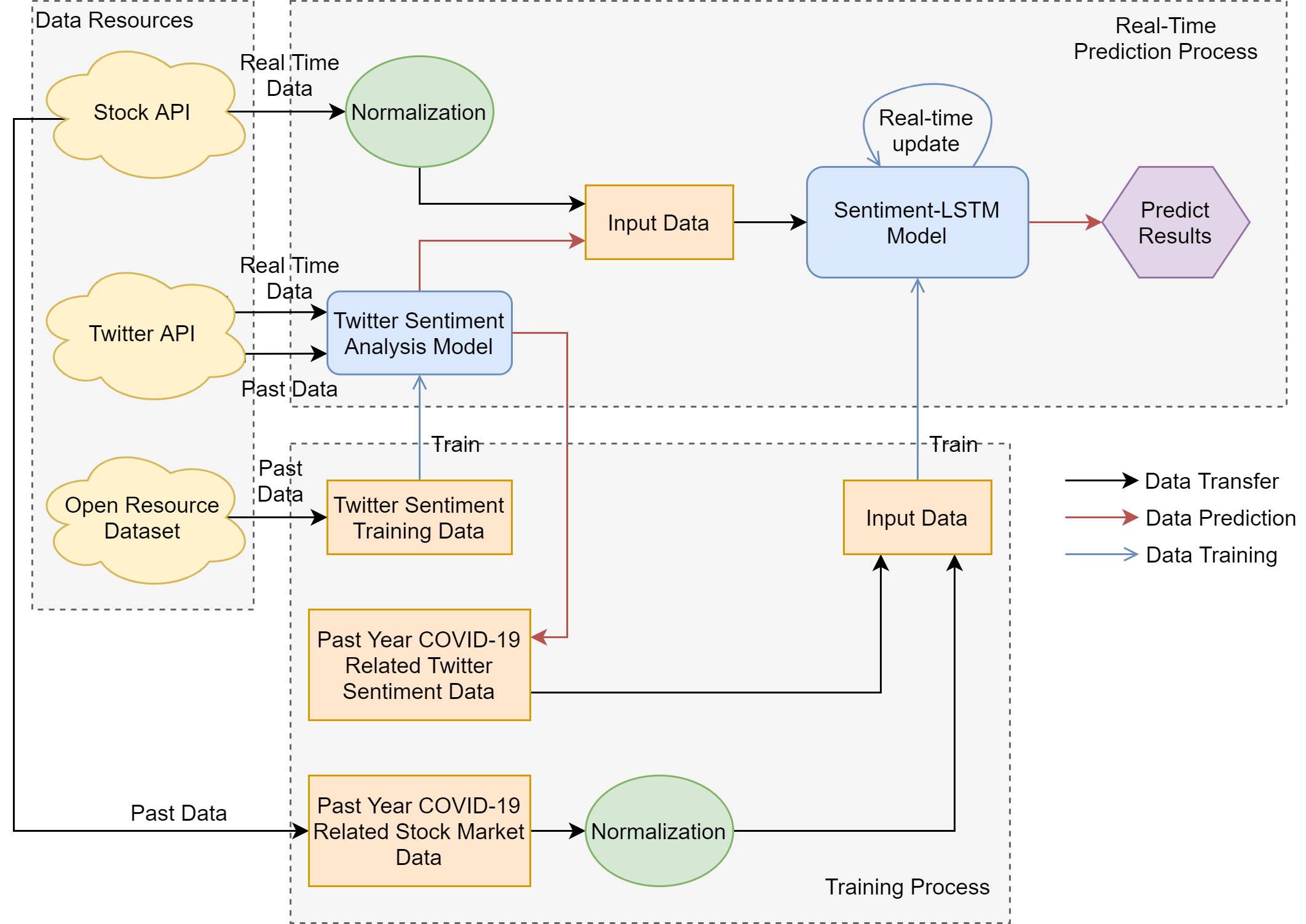}
    \caption{The Fundamental Process Flowchart of the Proposed Framework}
    \label{fig:system_flowchart}
\end{figure}

The fundamental process flowchart of the proposed framework is shown in Figure \ref{fig:system_flowchart}. The whole process can be divided into two major parts, which are Real-Time Prediction Process and Training Process. For the Real-Time Prediction Process, the data resources are Stock API and Twitter API, which can output real-time data for our proposed framework. The Stock API data is normalized as mentioned in Section 3.3 and combined with the Twitter Sentiment Analysis Model to be used as input data for the Sentiment-LSTM Model. For Training Process, the proposed framework uses an Open Resources Dataset to get Twitter sentiment training data then train the Twitter Sentiment Analysis Model. The Sentiment-LSTM Model in our proposed framework can self-update with the real-time data. It can help the proposed model fit the current situation better than a normal LSTM model.

\subsection{Time Series Model}

In this paper, we use traditional financial time series of neural networks~\cite{chen2016financial} and the information propagation formula can be written as
\begin{eqnarray}
    f_t &=& \sigma (W_{if}(x_t, s_t)+b_{if}+W_{hf}h_{i-1}+b_{hf}) \\
    i_t &=& \sigma (W_{ii}(x_t, s_t)+b_{ii}+W_{hi}h_{t-1}+b_{hi}) \\
    o_t &=& \sigma (W_{io}(x_t, s_t)+b_{io}+W_{ho}h_{t-1}+b_{ho}) \\
    g_t &=& \tanh (W_{ig}(x_t, s_t)+b_{ig}+W_{hg}h_{t-1}+b_{hg}) \\
    c_t &=& f_t \odot c_{t-1} + i_t \odot  g_t \\
    h_t &=& o_t \odot \tanh c_t
\end{eqnarray}
where $f,i,o$ represents the proportionality coefficients of forgetting, input, and output, respectively, and $g,c,h$ represents the candidate state, cell state, and hidden layer state, respectively. The scale coefficients are all used with a sigmoid function to limit the range of coefficients, and the candidate state is related to the information of the input and the hidden state of the previous time layer. The cell state can be considered as a kind of memory cell, and when updating the memory cell, the previous memory is selected to be partially forgotten and the new information is partially accepted, and the hidden layer values get information directly from the current memory cell state into a valve output (the output coefficient $o_t$).

\subsection{Detail Architecture of S-LSTM Model}

\begin{figure}[h]
    \centering
    \includegraphics[width=1\textwidth]{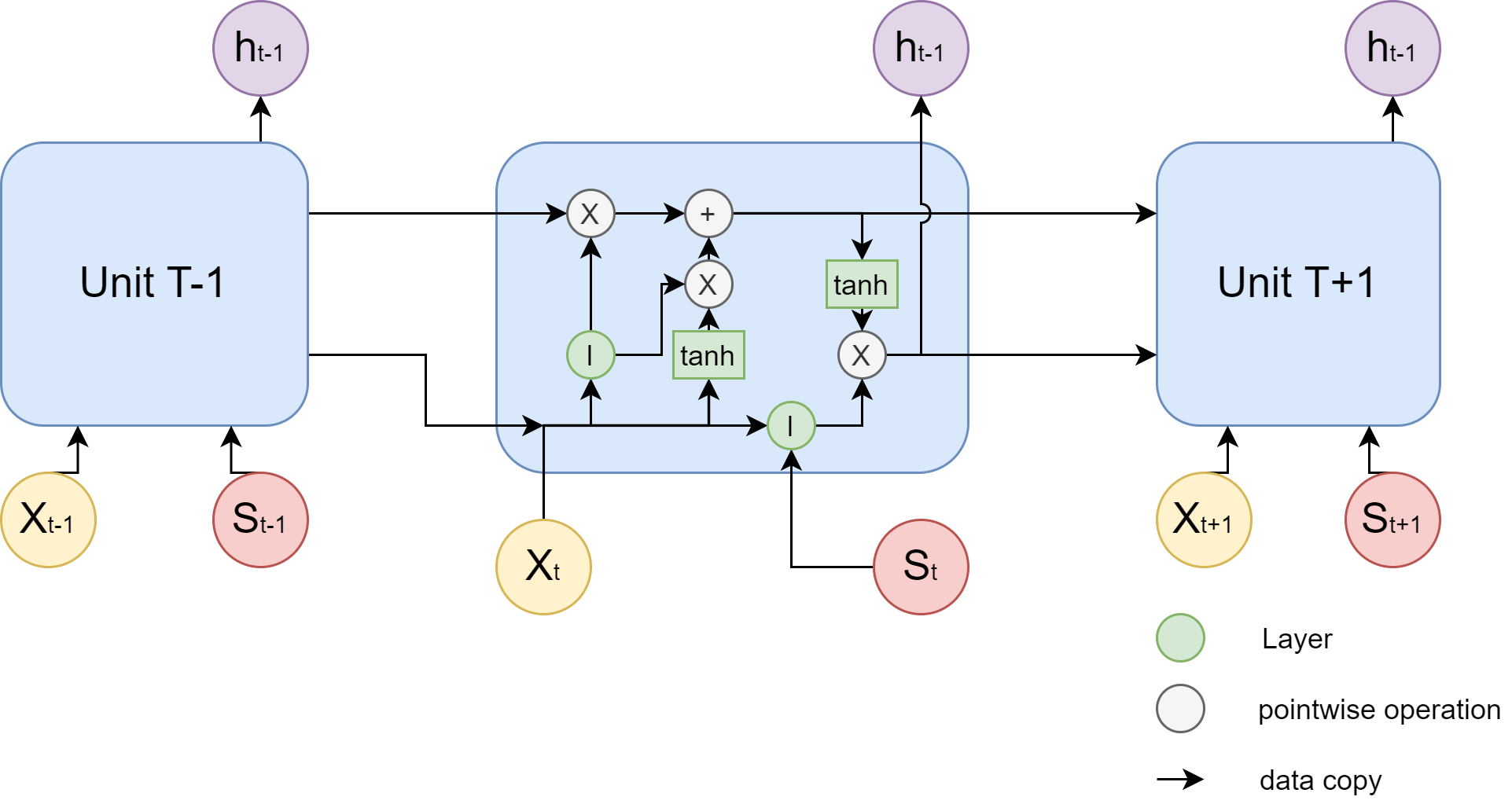}
    \caption{The Diagram of Unit T in the Proposed LSTM-S Algorithm.}
    \label{fig:LSTM-S}
\end{figure}

Based on the standard time-series LSTM model, we designed the Sentiment-LSTM model. This model incorporates the Sentiment Score we generated using the sentiment analysis model as the input data and combines the historical prices of stocks as another part of the input data. The specific algorithm unit can be seen in Figure \ref{fig:LSTM-S}. For each unit, in addition to reading the inputs $X_t$ and $S_t$ at the time point, the results generated from the previous time series are read from Unit $T-1$ as the input at this time point. Unlike traditional LSTM, we read the matrix of the Sentiment Score together with the time points of the algorithm unit and extend the input layers to absorb both stock prices and Sentiment Score and perform learning and regression.

In the diagram, each line transmits an entire vector, from the output of one node to the input of the other nodes. The gray circles represent pointwise operations, such as the sum of vectors, while the green matrix is the learned neural network layer. Lines that are joined together indicate the connection of vectors, and lines that are separated indicate that the content is copied and distributed to different locations.

\begin{figure}
    \centering
    \includegraphics[width=1\textwidth]{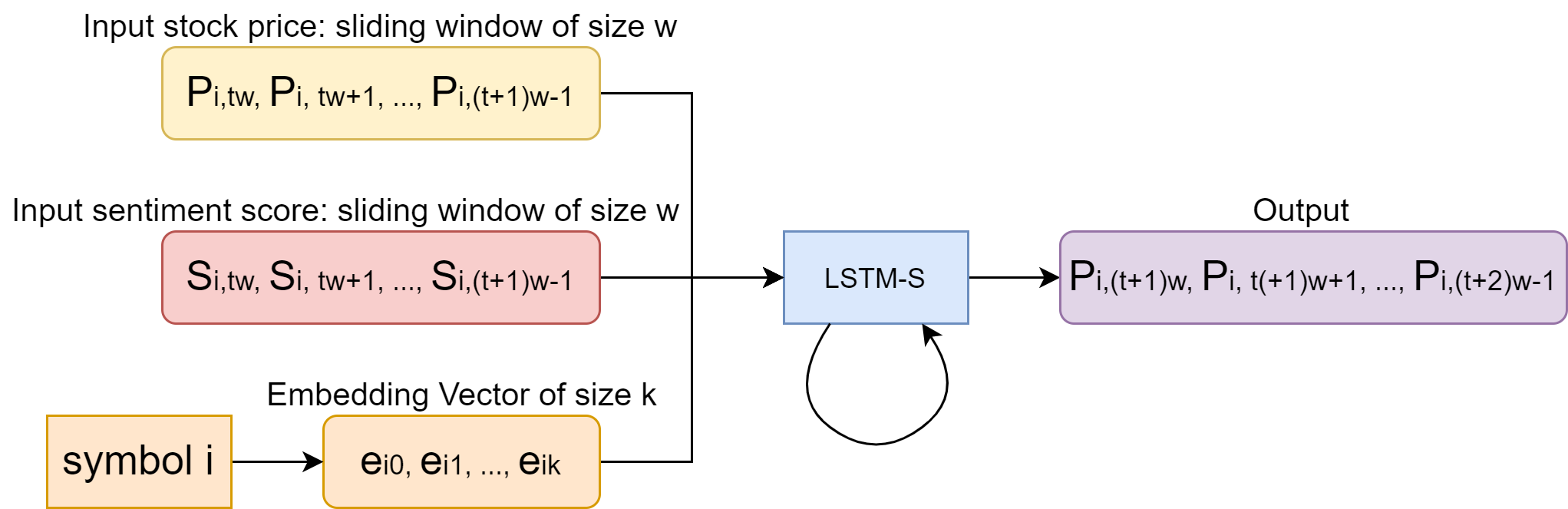}
    \caption{The proposed Sentiment-LSTM stock performance mining framework ($w=Size_{input}$, $k=Size_{embedding}$, and symbol $i$ can be any stock or stock index)}
    \label{fig:Framework}
\end{figure}

The Figure \ref{fig:Framework} explains in more detail how our model handles the data with input and output matrix. From the figure we can see that the input matrix consists mainly of two matrices, stock price and sentiment score at time $t$. The output is the stock price at time $t+1$.

\section{Experiment}

All the testing algorithms and proposed algorithm are implemented on PowerEdge C4130 Server with $2\times$ Intel Xeon E5-2690 v4 2.6GHz, 35M Cache, 9.6GT/s 1 QPI, Turbo, HT, 14C/28T, 512GB RAM, $4\times$ NVIDIA Tesla P100 16GB Passive GPU. The platform is PyTorch 1.4 and Tensorflow 2.0, the CUDA version is 10.1.

The pre-trained models are used to predict the benchmarks of the stock price changes during COVID-19. We use partial data before the September-October 2020 U.S. election as a test set to avoid the dramatic impact of other major events on stock prices. The Table~\ref{tab:stand} shows the performance of the major stock market forecasting models on stock price forecasts for the COVID-19 period. As expected, existing deep neural network stock market prediction models are unable to effectively predict stock market prices when major events occur. This is because these prediction models are trained based on data from the past decades and are fitted based on the patterns of those decades and cannot cope with sudden major events or changes. We can find that under normal conditions, some existing prediction models such as TCN, CNN, and LSTM can obtain an accuracy rate greater than $50\%$ and an F1-Score higher than $0.5$. However, when testing the trained stock market prediction models with the data during the epidemic, only the LSTM obtained slightly higher than $50\%$ accuracy and none of the tested models could obtain an F1-Score higher than 0.5 for the epidemic test data. Also, every tested prediction model showed more than a $4\%$ drop in accuracy when predicting the test data during COVID-19.

\begin{table}[]
\caption{Performance of the main forecasting models in the DJIA data set}
\begin{tabular}{l|lllll}
\hline
Model & Accuracy(Normal) & F1-Score(Normal) & Accuracy(Covid-19) & F1-Score(Covid-19) & Difference \\ \hline
ARIMA & 48.16\%          & 0.345            & 43.40\%            & 0.298              & -4.76\%    \\
TCN   & 56.48\%          & 0.514            & 48.69\%            & 0.441              & -7.79\%    \\
CNN   & 53.28\%          & 0.522            & 49.13\%            & 0.473              & -4.15\%    \\
LSTM  & 55.71\%          & 0.508            & 51.21\%            & 0.451              & -4.50\%   
\end{tabular}
\label{tab:stand}
\end{table}

\begin{table}[]
\caption{Input Data for Different Models}
\begin{tabular}{l|llllll}
\hline
Model                   & ARIMA & CNN & TCN & WB-TCN         & LSTM & S-LSTM              \\ \hline
Raw Data                & R     & R   & R   & N              & R    & S + R               \\ 
Processed Training Data & R     & R   & R   & word embedding & R    & sentiment score + R 
\end{tabular}
\label{tab:Input_data}
\end{table}

The algorithm we use to uncover the relationship between the stock market and social media sentiment is Sentiment(S)-LSTM, which, as mentioned before, can combine past stock market data to extract time-series patterns and combine them with the social media sentiment scores we obtain from our sentiment analysis model to learn and obtain a stock market index prediction model. The input data types for the different tested model are shown in Table~\ref{tab:Input_data}, where 1) Time series stock price data R: stock price dataset consisting of daily records of the Dow Jones Industrial Average; 2) Text news data N: news dataset consisting of historical news from the Reddit WorldNews channel; 3) Twitter sentiment analysis data S: sentiment scores generated based on relevant Twitter data.

\begin{figure}[h]
    \centering
    \includegraphics[width=1\textwidth]{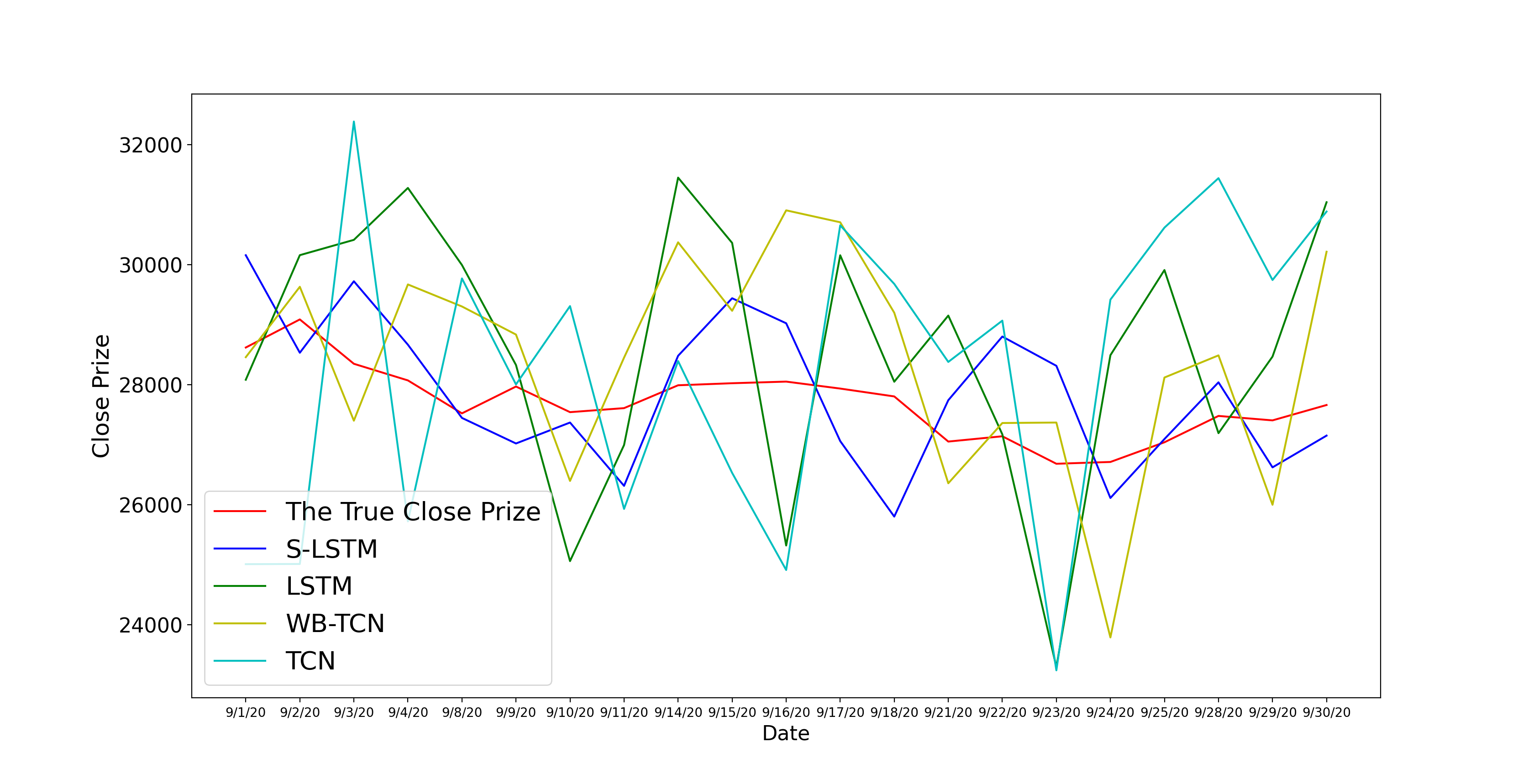}
    \caption{ Sep 2020 stock Dow Jones Industrial Index close price prediction comparison}
    \label{fig:predict}
\end{figure}

We tested different models using the Dow Jones Industrial Index Close Prize for each day of September 2020 as a test dataset. The input data are the Dow Jones Industrial Index Close Prize for the previous three days, the Twitter sentiment scores for the previous three days, and the COVID-19 related news text data for the previous three days. This is due to the relative stability of the U.S. stock market in September 2020, as well as the absence of events of great impact both domestically and internationally in the United States. Compared to November, the U.S. stock market is more volatile due to the election, which is not conducive to testing the performance of different models relative to COVID-19. The results of the test are shown in Figure~\ref{fig:predict}.

We use equation $|\frac{P_{predict} - P_{truth}}{P_{truth}}|$ to calculate the accuracy of different tested models and shown in Table~\ref{tab:predict}, the high percentage means that the model can obtain the correct trend (up or down) when predicting the stock price at the next point in time. In this test, we can clearly see that S-LSTM is the most accurate prediction algorithm, with a significant improvement compared to the traditional LSTM. Meanwhile, WB-TCN, which combines news text output, also shows advantages over traditional TCN and LSTM. This proves that the model combining more outputs has unparalleled advantages in dealing with unexpected events.

\begin{table}[h]
\centering
\caption{Input Data for Different Models}
\begin{tabular}{l|llll}
\hline
Model                   & TCN & WB-TCN         & LSTM & \textbf{S-LSTM}              \\ \hline
Accuracy                & 53.37\%   & 63.74\%   & 60.01\%    & \textbf{70.52}\%               \\ 
\end{tabular}
\label{tab:predict}
\end{table}

In order to compare the performance of the proposed algorithm for different single stock price predictions, we selected three stocks from different industries that are highly affected by COVID-19, AMC, CCL, and PFE, for our experiments. 
Like other companies in the industry, AMC Theatres has been negatively impacted by the coronavirus pandemic. As a result of the outbreak, AMC Theatres was forced to close hundreds of theaters. Upon reopening, they experienced low customer traffic as customers preferred to stay home to watch movies over watching movies at the theatres. Hollywood blockbusters were delayed due to COVID-19 as well. In October 2020, AMC Theatres warned investors that its dwindling cash reserves could force it to file for bankruptcy protection. Before the outbreak, the company was already \$4.75 billion in debt. Carnival Cruise Lines (CCL) is one of the top three cruise lines in the world. As of now, there is a lot of uncertainty about the cruise industry's return to normal operations. All three of these major cruise lines ended their 2020 fiscal year with record losses due to the New Crown Pneumonia outbreak. Carnival Corporation reported a net loss of \$10.236 billion on revenues of \$5.595 billion. These results reflect the enormous impact of the New Crown Pneumonia outbreak on the entire cruise industry. Pfizer (PFE) was the first major pharmaceutical company to announce the effectiveness of its new crown vaccine. So far this year, Pfizer's stock price has fallen before Moderna Inc. (MRNA) took a hit in November 2020 by announcing the effectiveness of its competing vaccine. In 2020, pharmaceutical stocks under-performed the rest of the class in the U.S. stock market, not an obvious outcome against the backdrop of a new crown epidemic spurring drug development. Even biotech stocks, of which Moderna is a part, have under-performed the S\&P 500.

The training data is the closing price of these three stocks during February 2020 to July 2020. The testing data is the closing price of these three stocks in August 2020 and September 2020. The models will use the closing stock price and sentiment score of the previous days as inputs to predict the closing stock price of the following day. We calculate the performance (lower means better performance) of the proposed S-LSTM algorithm by the following equation:
\begin{eqnarray}
    Diff_{t} &=& \frac{|P_{true,t} - P_{prediction,t}|}{P_{true,t}} \\
    Performance_{stock} &=& \sqrt{\frac{\sum_{t=1}^{n}Diff_{t}^2}{n-1}}
\end{eqnarray}
and the results are shown in Table \ref{std_stocks}. We can find that among the three stocks, the performance of our proposed S-LSTM is consistently the best. In comparison, PFE achieves relatively good performance on this stock for all four algorithms because of its small variation. For the two stocks, AMC and CCL, investors are more sensitive to the changes of the epidemic, so our proposed S-LSTM algorithm and the WB-TCN algorithm, when used as a comparison, appear to be more accurate for price prediction due to the combination of crowd sentiment scores for COVID-19 on social media or keyword extraction for COVID-19's news vocabulary. Specifically, our proposed algorithm incorporating sentiment scores provides better prediction results than previous WB-TCNs.

\begin{table}
\centering
\caption{Performance of stock closing price prediction results for the four algorithms for the three stocks during August and September 2020}
\begin{tabular}{l|llll}
\hline
Stock & \textbf{S-LSTM}                      & LSTM                        & WB-TCN                      & TCN                         \\ \hline
AMC   & \textbf{0.11472}                     & 0.24194                     & 0.17614                     & 0.25891                     \\
CCL   & \textbf{0.12040}                     & 0.23028                     & 0.173659                    & 0.26957                     \\
PFE   & \textbf{0.11525} & 0.17657 & 0.14922 & 0.21379
\end{tabular}
\label{std_stocks}
\end{table}

\begin{figure}
    \captionsetup[subfloat]{justification=centering}
    \centering
    \subfloat[Stock AMC Entertainment (AMC) Close Prise Prediction]{\includegraphics[width=1\textwidth]{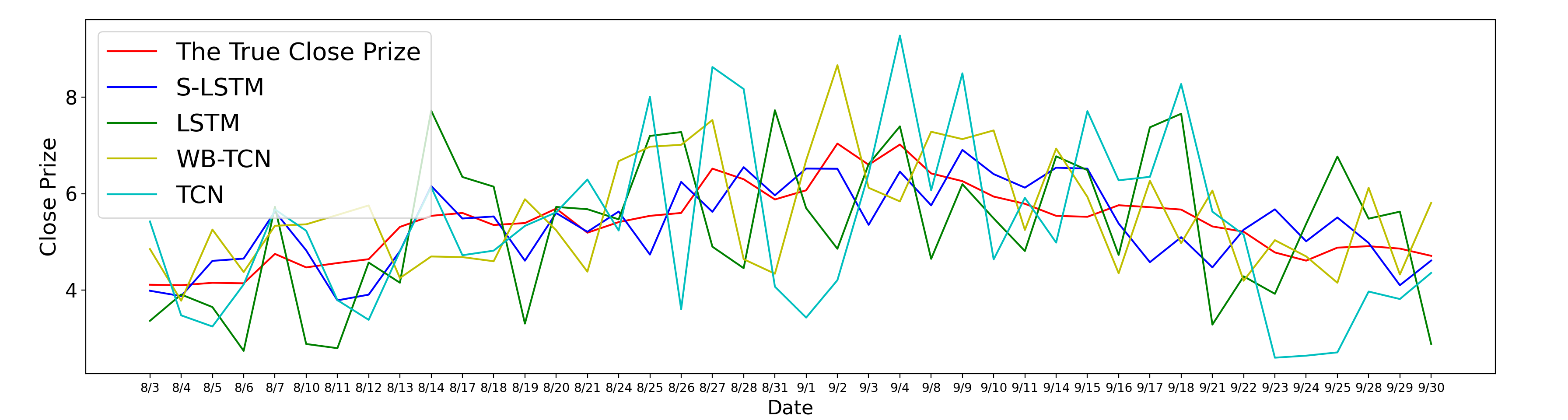}} \\
    \subfloat[Stock Carnival Corp (CCL) Close Prise Prediction]{\includegraphics[width=1\textwidth]{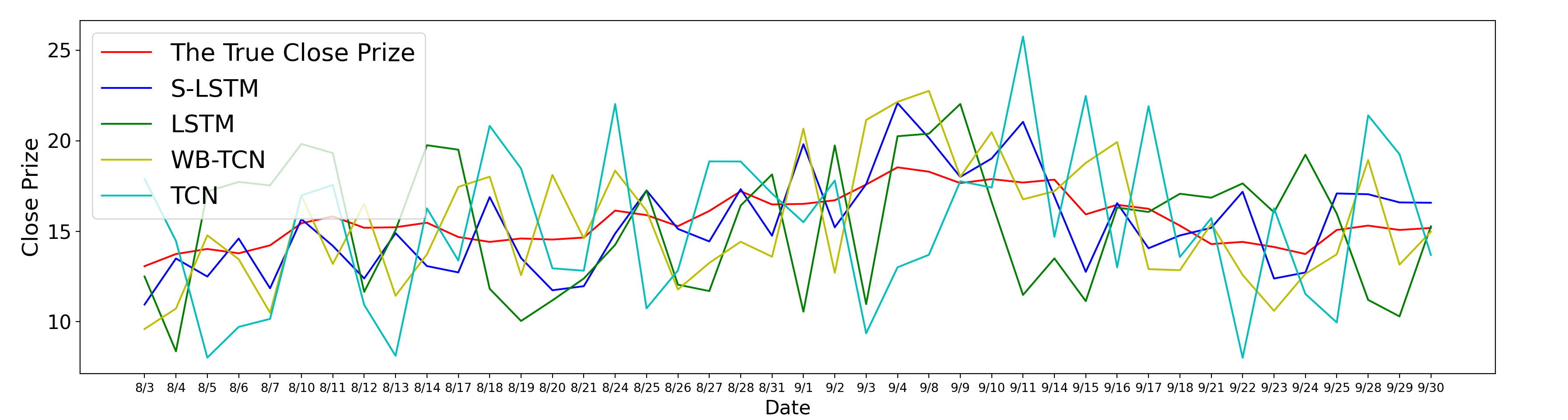}} \\
    \subfloat[Stock Pfizer Inc. (PFE) Close Prise Prediction]{\includegraphics[width=1\textwidth]{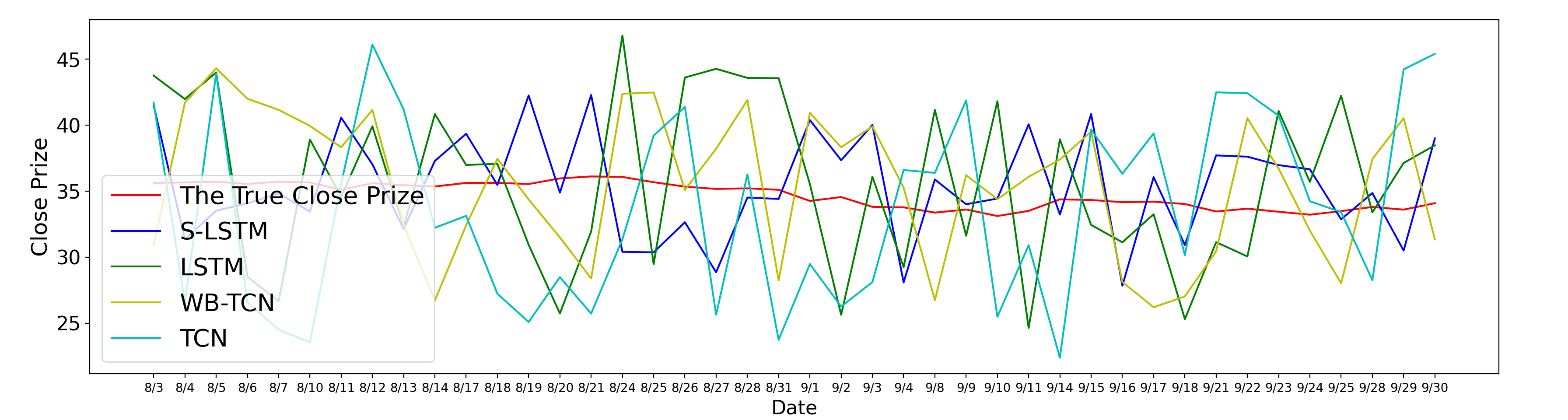}}
    \caption{August and September 2020 stock close price prediction comparison}
    \label{fig:stock_prediction}
\end{figure}

We also plot out the forecast results to show in the Figure \ref{fig:stock_prediction}, and we can see that among the three stocks, PFE stock has relatively the least change. All algorithms on PFE zoom in on the changes and appear to be not very accurate. CCL and AMC are more sensitive to changes in the COVID-19 outbreak due to the large changes and the properties of the stocks themselves, so the closing prices of the stocks predicted by our proposed algorithm are more in line with the actual trend.

\section{Discussion}
As we have shown in Table 2, LSTM neural network may not be enough to predict stock market price. It is combined with text sentiment over a short period of time in our experiment, as seen in Section 4, that we can make full use of this model.  Based on the results, our implicates that when sentiment is applied to neural networks, most notably the S-LSTM model, it will yield better results and can be used as a model for closing price of stocks to a high degree of accuracy.

Quantitative analysis work can be divided into three parts from a broad perspective: macro, meso and micro. Macro, through the study of macroeconomic data to analyze and judge the future economic trends, meso is based on industry data, industry trends, rotation, etc., while micro is based on the company's fundamental data, stock selection.

Economic trends will largely be reflected in the stock market, the economic form is good, the stock market for the probability of bull market. When the economic fundamentals are weak and downward pressure is high, the stock market is lukewarm. For example, in 2018, the overall economic downturn, the stock market fell more, and stock market trends and macroeconomic trends show a strong correlation. Macroeconomic forecasts will be the most important reference when making investment decisions, and industry and firm-level analysis will be done on this basis.

Existing models basically use stock market data from the past decades for bench marking, to measure the general market trend, and to forecast future trends based on macro data analysis. However, the strategies among them take into account fewer factors and are unlikely to explain the complexity of the stock market, and need to be used in conjunction with other strategies.

From the performance of different models on the test set of stock market data during COVID-19, we can see that the event sequence has a clear deficiency for complex stock market prediction. Even the published stock market forecasting models are not able to predict stock market fluctuations well when encountering large unexpected events. In this case, a prediction model that contains more information can be more advantageous. Whether it is news information or social media information, it allows the model to obtain more macro information and make more reliable forecasts.

In terms of specific model inputs, the emotional responses of users on Twitter caused by COVID-19 can significantly improve the prediction accuracy of the LSTM model. Compared with the news text output for TCN, the improvement of Twitter sentiment analysis is greater.

\section{Conclusion}
We want to understand if we can further the research of using text sentiment from social media as news to predict stock market prices. With social media as a source of news, we use text sentiment to further the econometric domain in understanding and predicting stock market prices. We accomplish the highest model accuracy through an S-LSTM model. We highlight the data processing required for text sentiment analysis as well as financial time series data on market indexes performances over time. With a different fit for each index, we measured the performance of ARIMA, TCN, CNN, and LSTM models with DJIA data set. Lastly, we ran an experiment on the time periods of September 2020 - October 2020 pre-trained to discover than it was the S-LSTM gave us the best fit. We found a range of correlation of COVID-19 sentiment with prices. In conclusion, further research of applying text sentiment to predict stock market performance is necessary. The more insight that text sentiment data can give us, the more news or information we are able to apply to reach full information and reflect the real value of the stock price.





\printendnotes

\bibliography{main}



\end{document}